\documentclass[twocolumn]{webofc}

\usepackage[varg]{txfonts}   
\usepackage{hyperref}
\usepackage{amsmath}
\usepackage{amssymb}
\usepackage{graphicx}
\usepackage{float} 

\newcommand{\qgsjet}{\textsc{QGSJet}}

\newcommand{\einv}{E_{\mathrm{inv}}}
\newcommand{\ecal}{E_{\mathrm{cal}}}
\begin{document}
\title{Determination of the invisible energy of extensive air showers from the data collected at Pierre Auger Observatory}
%
%

\author{\firstname{Analisa G.} \lastname{Mariazzi}\inst{1}\fnsep\thanks{\email{mariazzi@fisica.unlp.edu.ar}} 
        \firstname{for the } \lastname{Pierre Auger Collaboration }\inst{2,3}\fnsep\thanks{\email{
        auger\_spokespersons@fnal.gov}} 
}

\institute{
Instituto de F\'isica La Plata, CONICET, La Plata, Argentina 
\and
Observatorio Pierre Auger, Av. San Martin Norte 304, Malarg\"ue, Argentina 
\and
Full author list: \url{http://www.auger.org/archive/authors_2018_10.html}
          }

\abstract{In order to get the primary energy of cosmic rays from their extensive air showers using the fluorescence detection technique, the invisible energy should be added to the measured calorimetric energy. The invisible energy is the energy carried away by particles that do not deposit all their energy in the atmosphere.
It has traditionally been calculated using Monte Carlo simulations that are dependent on the assumed primary particle mass and on model predictions for neutrino and muon production.

In this work the invisible energy is obtained directly from events detected by the Pierre Auger Observatory. The method applied is based on the correlation of the measurements of the muon number at the ground with the invisible energy of the showers. By using it, the systematic uncertainties related to the unknown mass composition and to the high energy hadronic interaction models are significantly reduced, improving in this way the estimation of the energy scale of the Observatory.}
\maketitle
\section{Introduction}
\label{intro}
Above $10^{15}$ eV cosmic rays are detected indirectly through the extensive air showers (EAS) they produce in the atmosphere. Most of the cosmic ray energy is carried by electromagnetic particles of the EAS, which can be detected by their secondary electromagnetic signatures, e.g. radio, Cherenkov or fluorescence light. 
In the case of the fluorescence detection, the fluorescence radiation emitted by the nitrogen molecules of air excited by the charged particles of the EAS is produced in proportion to the energy dissipation, allowing a reconstruction of the longitudinal profile of the energy deposit ($dE/dX$) of the shower as a function of the atmospheric depth $X$.
The atmosphere is used as a calorimeter and the integral $\int \left(dE/dX\right) dX$, called the calorimetric energy of the shower, $E_{\rm cal}$, is measured. 

$E_{\rm cal}$ underestimates the total shower energy ($E_0$) because neutrinos do not suffer electromagnetic interactions and high energy muons reach ground level after releasing only a portion of their energy into the atmosphere. Thus, an estimation of the primary energy $E_0$ with the fluorescence detection technique is obtained by adding to $E_{\rm cal}$ a correction to account for the {\it invisible} energy ($E_{\rm inv}$) carried by the particles that do not dissipate all their energy in the atmosphere. 
$E_{\rm inv}$ amounts to about 10\% - 20\% of $E_0$.
\begin{figure*}[t]
\centering
 \includegraphics[width=1\columnwidth,height=1\columnwidth,keepaspectratio,clip] {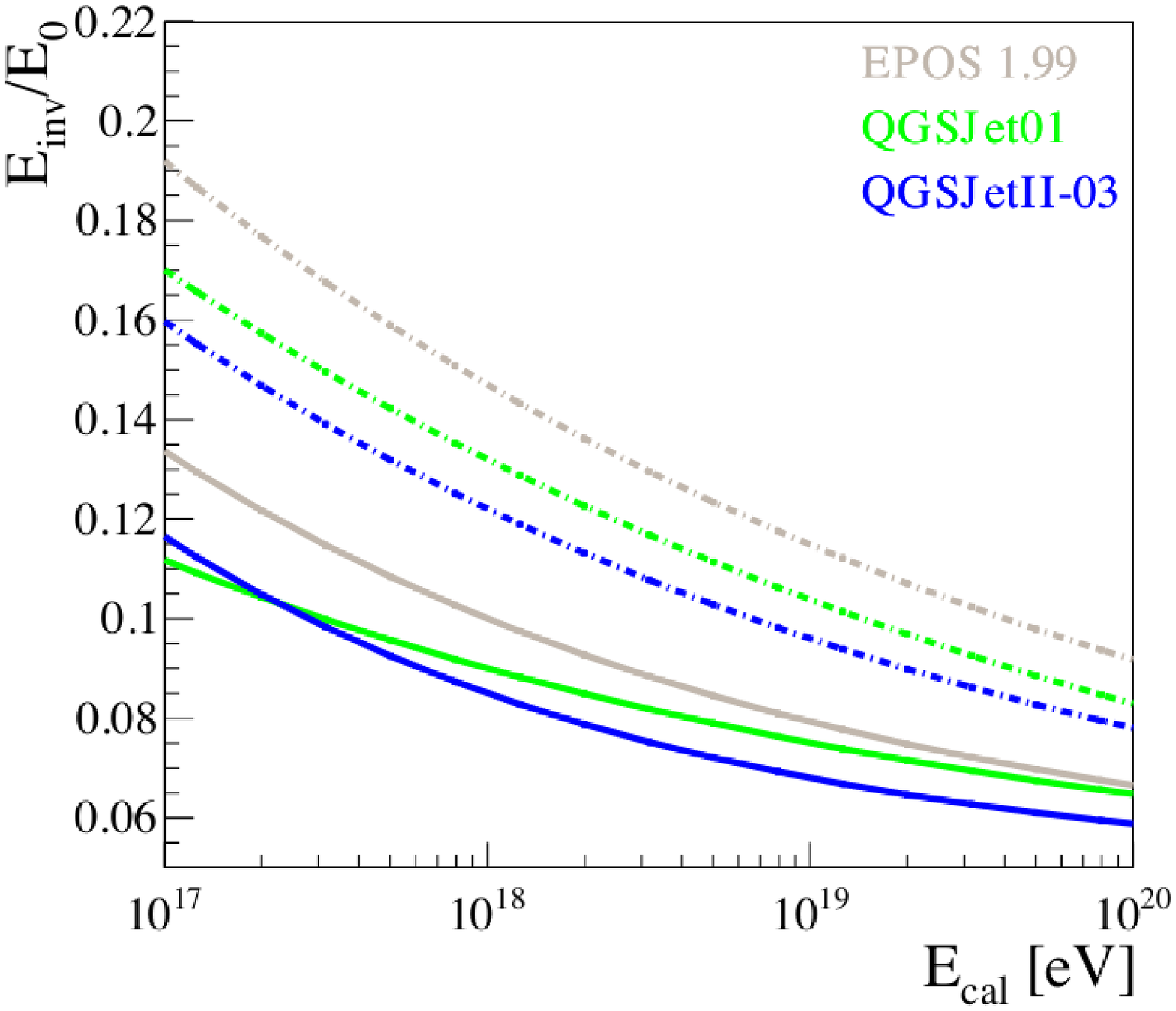}
 \includegraphics[width=1\columnwidth,height=1\columnwidth,keepaspectratio,clip] {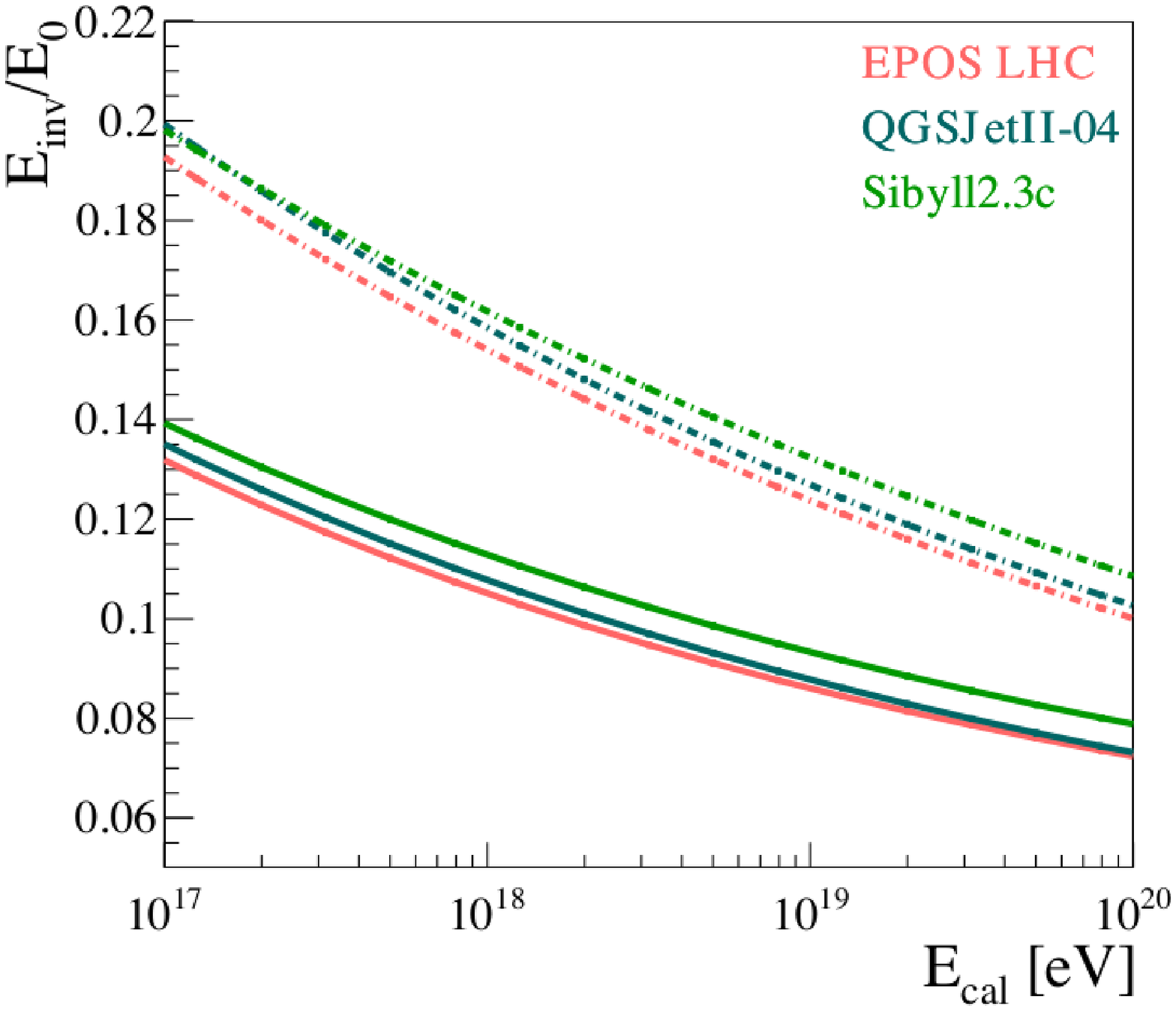}
\caption{Average invisible energy fraction as a function of  $E_{\rm cal}$ calculated with Monte Carlo simulations using the hadronic interaction models tuned with LHC data (right) and the models developed before the LHC data were available (left). The predictions for proton and iron primaries are shown with solid and dashed lines, respectively.
The simulations were performed using the CORSIKA code for the models EPOS 1.99 and \qgsjet II-03~\cite[\& references therein]{CORSIKA} and with the AIRES code for \qgsjet 01~\cite[\& references therein]{AIRES}. For the models tuned with the LHC data EPOS LHC, \qgsjet II-04, and Sibyll2.3c we used the CONEX code~\cite[\& references therein]{CONEX}.
}
\label{fig:Einv-MC}      
\end{figure*}

$E_{\rm inv}$ can be calculated directly from the energy deposited in the atmosphere by the different components of simulated air showers~\cite{Barbosa}. 
There are large differences in the values of the ratio of $E_{\rm inv}$ to $E_0$ as a function of $\ecal$ for different hadronic models and primary masses, as could be seen in Fig.~\ref{fig:Einv-MC}.
The estimation of $E_{\rm inv}$ is affected by the irreducible uncertainties associated with the models describing the hadronic interactions and also by the mass composition of cosmic rays. 

The models to get $\einv$ can be improved further using the primary mass composition estimated with the fluorescence detectors~\cite{Auger-mass-comp} so that the spread between the predictions is significantly reduced for a given mass. 
However, the uncertainties associated with the hadronic interaction models are difficult to estimate and are ultimately unknown~\cite{Pierog-ICRC2017}. 
Even after the updates with LHC data, the models still fail to describe several properties of the shower development related to muons~\cite{Auger-HadInt-UHECR18}, and this can introduce unpredictable biases in the $\einv$ estimation. 

Thus the strategy followed in this work is to estimate $E_{\rm inv}$ using the correlations that exist between $\einv$ and shower observables measured at the Pierre Auger Observatory~\cite{Auger}, correlations that to a large extent are not sensitive to the hadronic interaction models and primary mass composition.  

The Pierre Auger Observatory~\cite{Auger} is a {\it hybrid} observatory, because the measurements 
are done combining the data of a Surface Detector (SD) that is sensitive to the muon content of EAS and a Fluorescence Detector (FD).
The SD consists of 1660 water-Cherenkov detectors (WCDs) arranged on a hexagonal grid of 1.5~km spacing extending over a total area of $\sim 3000~{\rm km}^2$. 
The FD consists of 24 telescopes placed in four sites located along the perimeter of the Observatory that overlook the atmosphere above the surface array. The FD operates during clear and moonless nights with a duty cycle of about 14\%~\cite{AugerFD}. 
\section{Phenomenology of the invisible energy}
\label{Sec:EinvPhenomenology}
The Heitler model and its extension to hadronic cascades~\cite{Heitler,Matthews} provide a qualitative description of EAS which is suitable enough to serve as a guiding thread in the next sections, where the starting points of the data-driven approaches to estimate $\einv$ will be inspired by some of the expressions outlined below.

In the model, only pions are produced in the hadronic interactions, all with the same energy
and  the same particle multiplicity ($N$). 
The neutral pions decay almost immediately into two photons, generating an electromagnetic cascade. 
Charged pions interact hadronically until the average energy of the charged pions is decreased to  
such a level that their time-dilated decay length 
becomes smaller than their hadronic interaction length 
. This energy is referred to as the pion critical energy: 
\begin{equation}
\epsilon^\pi_c = \frac{E_0}{N^{n}},
\label{eq:epi}
\end{equation}
where $E_0$ is the primary particle energy and $n$ is the number of interactions suffered by the charged pions. 

One important feature of the model is that $\einv$ is proportional to the number of muons ($N_\mu$) reaching ground level.
\begin{equation}
\einv = \epsilon^\pi_c ~N_\mu,
\label{eq:EinvNmu}
\end{equation}
This expression will be the guiding thread to estimate $\einv$ with a measurement of $N_\mu$ in inclined showers.

Another important feature of the model is the power law of $\einv$:
\begin{equation}
\einv=\epsilon^\pi_c\left(\frac{E_0}{\epsilon^\pi_c}\right)^\beta,
\label{eq:EinvE0}
\end{equation}
where $\beta = \ln(\frac{2}{3} N) / \ln N$. 
Air shower simulations predict values of $\beta$ in the range from 0.88 to 0.92~\cite{K-U-review}. 
$\beta$ also fixes the $\einv$ dependence on the mass number $A$ of the primary. 
Neglecting collective effects in the first interactions so that the cascade is the superposition of $A$ cascades initiated by primary protons of energy $E_0/A$ one has:
\begin{equation}
\einv ^A=\epsilon^\pi_c\left(\frac{E_0}{\epsilon^\pi_c}\right)^\beta A^{1-\beta}.
\label{eq:EinvE0A}
\end{equation}
This relationship will be the guiding thread to estimate $\einv$ 
from vertical showers measurements.

Monte Carlo simulations take into account  all the complex phenomena occurring throughout the EAS development giving more quantitative predictions of $\einv$, which is calculated following the method described in~\cite{Barbosa}.

\begin{figure}[t]
\centering
 \centerline{ 
    \includegraphics[width=0.98\columnwidth,height=0.88\columnwidth] {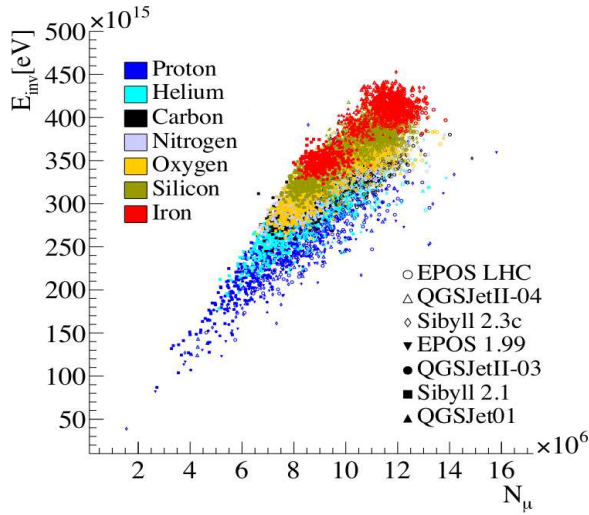}
 }
\caption{Correlation between $E_{\rm inv}$ and the number of muons reaching ground with energy greater than 100 MeV for different hadronic interaction models and primaries simulated with energy of $3 \times 10^{18}$ eV arriving at $60^\circ$ at the altitude of the observatory.}
\label{fig:Einv-Nmu}
\end{figure}

The correlation between $\einv$ and $N_\mu$ has been studied 
simulating showers with different primary masses and hadronic interactions models using the CORSIKA code~\cite[\& references therein]{CORSIKA}, as it is shown in Fig.~\ref{fig:Einv-Nmu}. 
In spite of the very large spread in the predictions of $N_\mu$ and $\einv$, the correlation is good and is similar for all models and primaries, suggesting that it is possible to obtain a robust estimation of $\einv$ from the measurements of $N_\mu$. 
\begin{figure*}[t]
\centering
 \centerline{
\includegraphics[width=1\columnwidth,height=0.9\columnwidth,clip]{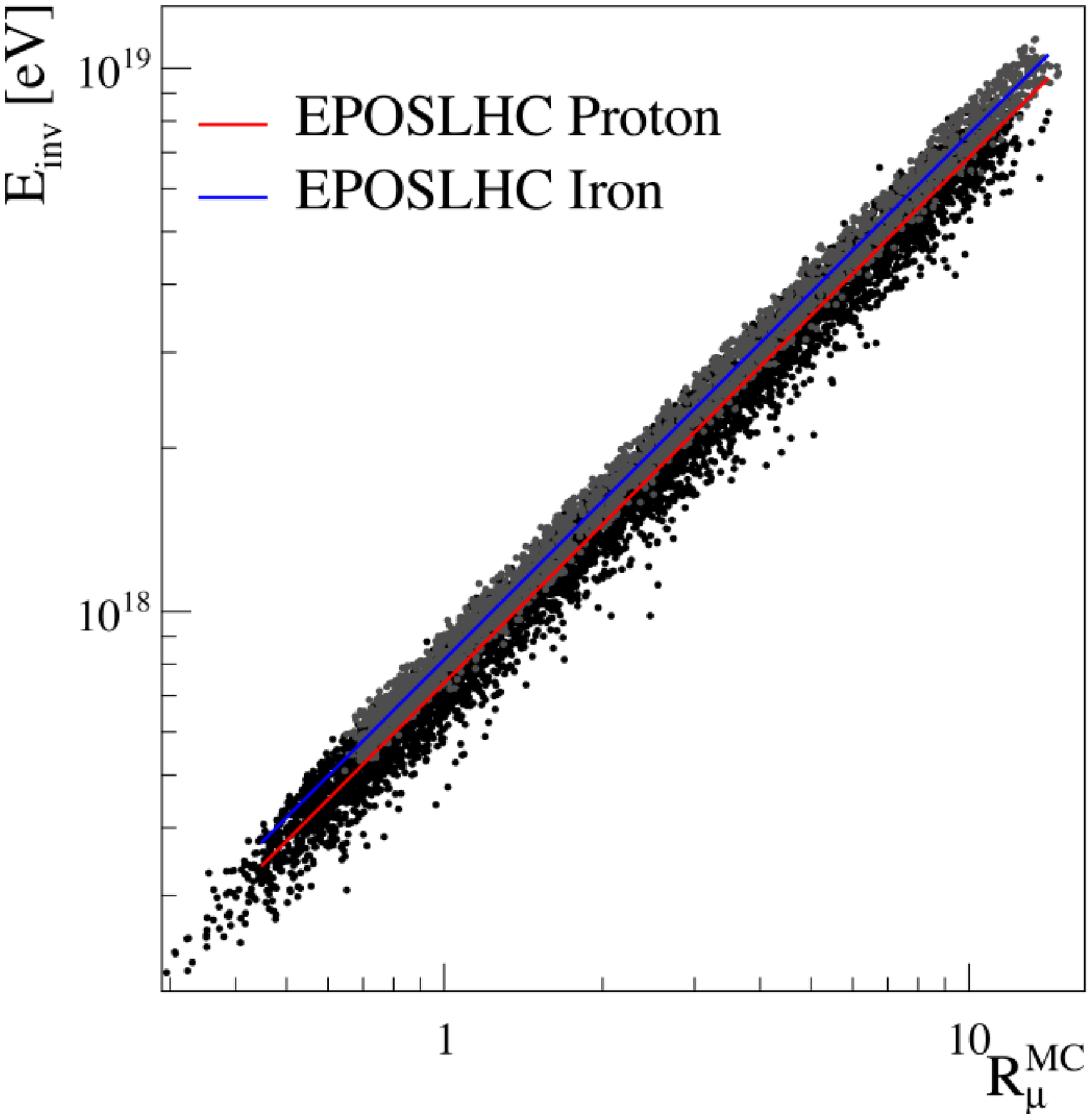}
\includegraphics[width=1\columnwidth,height=0.9\columnwidth,clip]{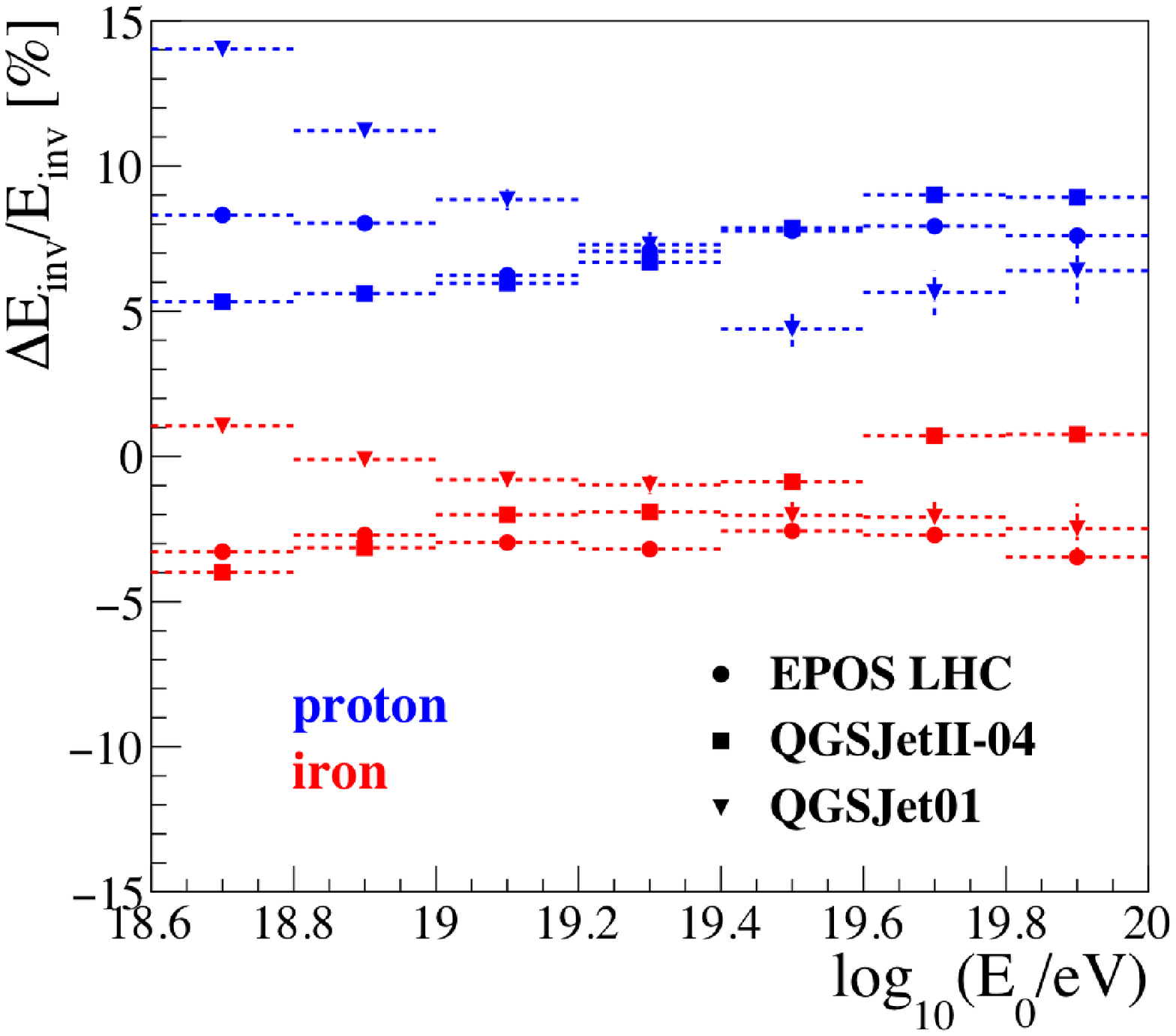}
}
\caption{Left: Correlation between $E_{\rm inv}$ and $R_{\mu}^{\rm MC}$ for the proton (black dots) and iron (grey dots) showers simulated with EPOS LHC. Right: Average value of the relative difference between the true value of $E_{\rm inv}$ and the value $E_{\rm inv}$ reconstructed using
the EPOS LHC parameterisation for a mixture of 50\% protons and 50\% iron.}
\label{fig:Einv_Rmu}
\end{figure*}
\begin{figure*}[t]
\centering
 \includegraphics[width=1\columnwidth,height=0.9\columnwidth,keepaspectratio,clip] {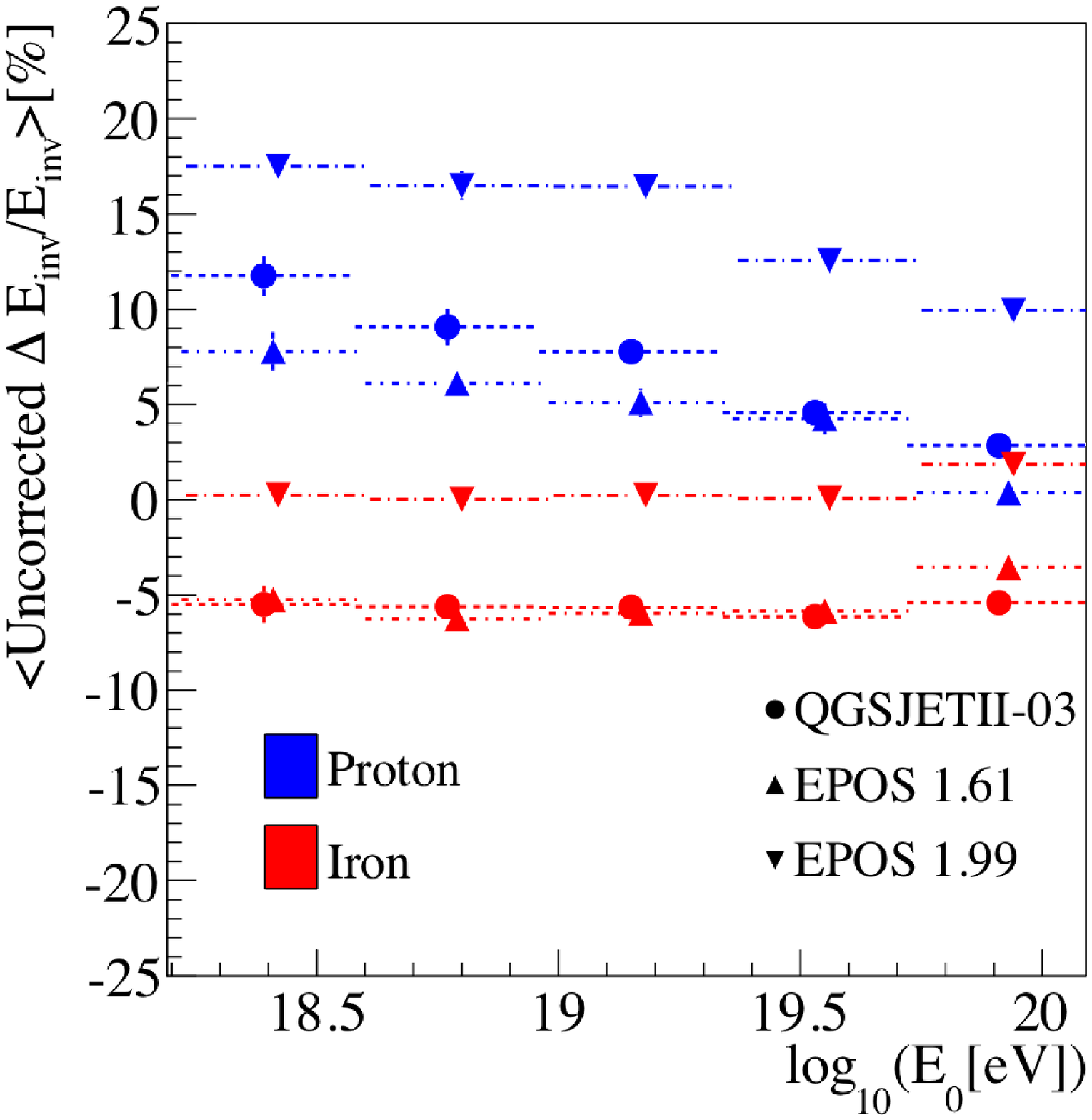}
 \includegraphics[width=1\columnwidth,height=0.9\columnwidth,keepaspectratio,clip] {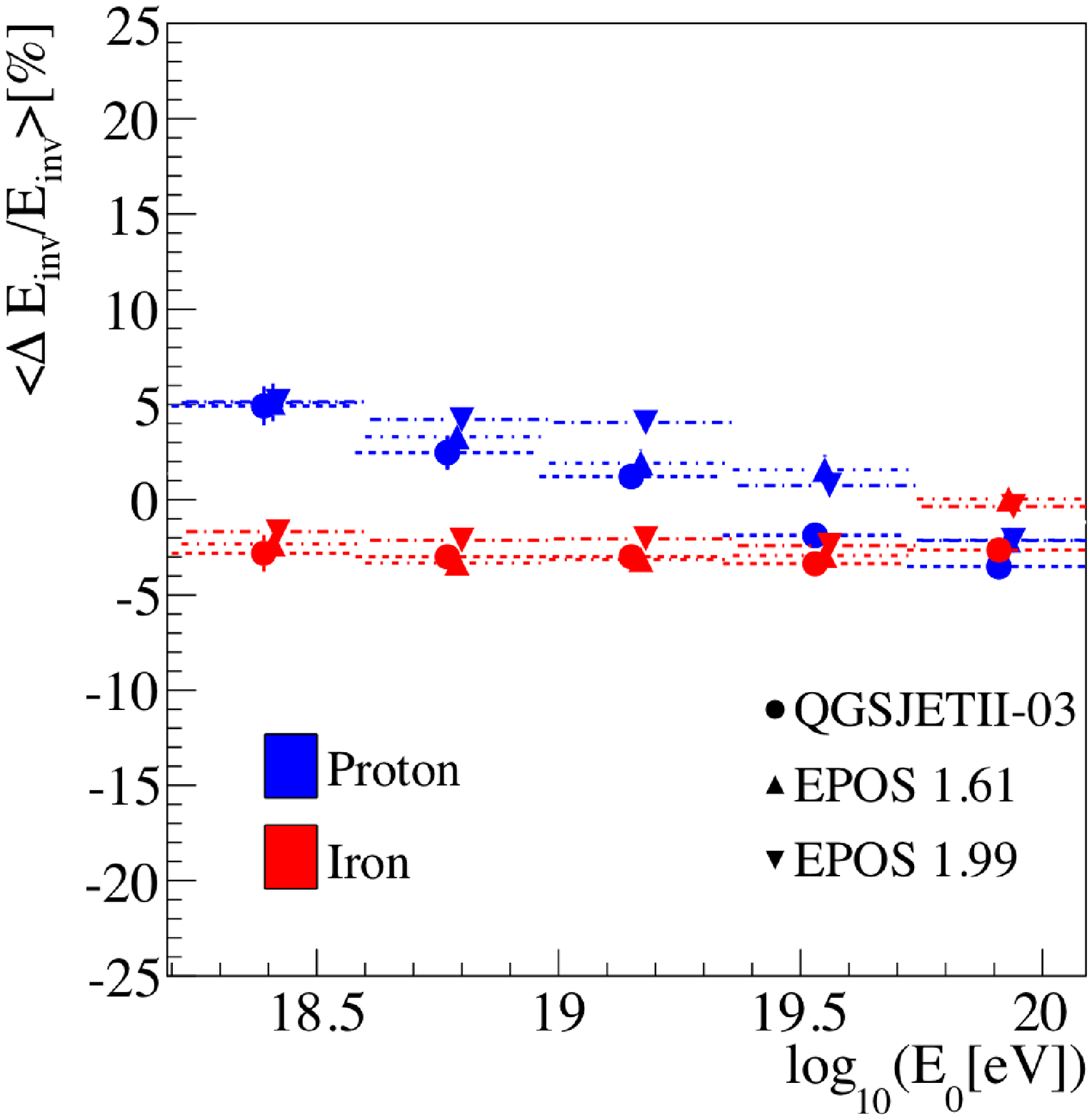}
\caption{Average value of the relative difference between the true value of $E_{\rm inv}$ and the value of $E_{\rm inv}$ reconstructed from $S(1000)$ and $X_{\rm max}$ using Eq.~\eqref{eq:Einv_S1000} 
before (right) and after (left) applying the correction due to the difference among the simulations in the predictions of the number of muons and in the attenuation function, using Eq.\eqref{eq:correction}.
        }
\label{fig:EinvS1000Bias}     
\end{figure*}
%
\section{Estimation of \texorpdfstring{$\einv$}{ei} using Auger data}
\label{Sec:EinvAnalysis}
Two different reconstruction techniques are used for the SD events: one for the so-called vertical showers with zenith angles $\theta < 60^\circ$~\cite{SDExp}, and one for the inclined showers with $\theta > 60^\circ$~\cite{HASRec}. 
WCDs are sensitive to the electromagnetic and hadronic components of a shower. 

The most straightforward way to estimate $E_{\rm inv}$ is to use inclined showers, in which the electromagnetic component of the shower is largely absorbed and it is possible to measure the total number of muons arriving at ground level which is an observable expected to be proportional to $E_{\rm inv}$, as seen in Sec.~\ref{Sec:EinvPhenomenology} (Eq.~\eqref{eq:EinvNmu}). 
The muon number cannot be directly measured for vertical events. However, $E_{\rm inv}$ can be obtained from the energy estimator using the power law relationship between $E_{\rm inv}$ and $E_0$ (see Eq.~\eqref{eq:EinvE0A}).
\subsection{\texorpdfstring{$E_{\rm inv}$} from inclined showers}
\label{Sec:Einv_Rmu}
The reconstruction of inclined events~\cite{HASRec} is based on the fact that the muon number distribution at ground level can be described by a density scaling factor that depends on $E_0$ and primary mass, and by a lateral shape that, for a given  arrival direction $(\theta,\phi)$ of the shower, is consistently reproduced by different hadronic interaction models and depends only weakly on $E_0$ and primary mass. 

The muon number density as a function of the position at ground $\vec{r}$ is then parameterised with
\begin{equation}
\rho_{\mu}(\vec{r}) = N_{19}~\rho_{\mu,19}(\vec{r};\theta,\phi),
\label{MDensity}
\end{equation}
where $\rho_{\mu,19}(\vec{r};\theta,\phi)$ is a reference distribution conventionally calculated for primary protons at $10^{19}$ eV using the hadronic interaction model \qgsjet II-03, and the scale factor $N_{19}$ represents the shower size relative to the normalization of the reference distribution.

The performance of the reconstruction is validated on simulated events. The reconstructed value of $N_{19}$ is compared with its true value $R_{\mu}^{\rm MC}$ for each simulated event.
$R_{\mu}^{\rm MC}$ is defined as the ratio of the total number of muons at ground level to the total number of muons in the reference model. The relative deviation of $N_{19}$ from $R_{\mu}^{\rm MC}$ is within 5\% for several hadronic interaction models and primaries~\cite{AugerMuonSize}. A bias correction is then applied to $N_{19}$ in order to reduce the residuals to within 3\% of the most recent models tuned with LHC data. In this way, the corrected value of $N_{19}$, which in the following is called $R_\mu$, represents an unbiased estimator of the total number of muons at ground level.

The correlation between $E_{\rm inv}$ and the total number of muons at ground level is studied with two data sets: one simulated with CORSIKA using the hadronic interaction models EPOS LHC and QGSJET II-04~\cite[\& references therein]{CORSIKA} 
and the other with AIRES using the model \qgsjet 01~\cite[\& references therein]{AIRES}.

For each simulated event, we calculate the values of $E_{\rm inv}$ and of the muon number at ground level $R_{\mu}^{\rm MC}$. 

For all the samples of simulated events, the correlation between $E_{\rm inv}$ and $R_{\mu}^{\rm MC}$ is well described by a power-law 
\begin{equation}
E_{\rm inv} =  C  ~  \left(R_{\mu}^{\rm MC}\right)^{\delta},
\label{eq:EinvN19}
\end{equation}
where the values of the parameters $C$ and $\delta$ are obtained from a fit to the events.
Examples of the correlation between $E_{\rm inv}$ and $R_{\mu}^{\rm MC}$ are shown for  in~Fig.~\ref{fig:Einv_Rmu} (left), where the lines show the fitted power law relationships. 

The relationship of Eq.~\eqref{eq:EinvN19} is used to estimate $E_{\rm inv}$ in the data from the measurement of $R_{\mu}$ that, as seen before, is the unbiased estimator of $R_{\mu}^{\rm MC}$. Since the mass composition of the data is not precisely known, the estimation of $E_{\rm inv}$ is obtained using the parameterisation of $E_{\rm inv}$ as a function of $R_{\mu}$ for a mixture of 50\% protons and 50\% iron. This is done taking the average of the two $E_{\rm inv}$ estimations in Fig.~\ref{fig:Einv_Rmu} (left) obtained for proton and iron primaries using the EPOS LHC hadronic interaction model.

The performance of the analysis is studied on fully simulated events\footnote{ events simulated with the detector response and $R_{\mu}$ reconstructed with the same algorithm used for the data}. 
For each event, we compute $E_{\rm inv}$ from $R_{\mu}$ using the estimation for the mixed proton and iron composition, and we compare it with the true value of $E_{\rm inv}$. 
The average values of the residuals as a function of the true value of $E_{\rm inv}$ are shown in Fig.~\ref{fig:Einv_Rmu} (right) for proton and iron primaries for EPOS LHC, QGSJET II-04 and \qgsjet 01 hadronic interaction models.
The residuals are within $\pm 10\%$ which is an indication of the overall systematic uncertainty in  $E_{\rm inv}$ estimation, which is dominated by the model and mass dependence of the values of $C$ and $\delta$.
\begin{figure*}[t]
\centering
 \centerline{
\includegraphics[width=0.9\columnwidth,height=0.9\columnwidth,keepaspectratio,clip]{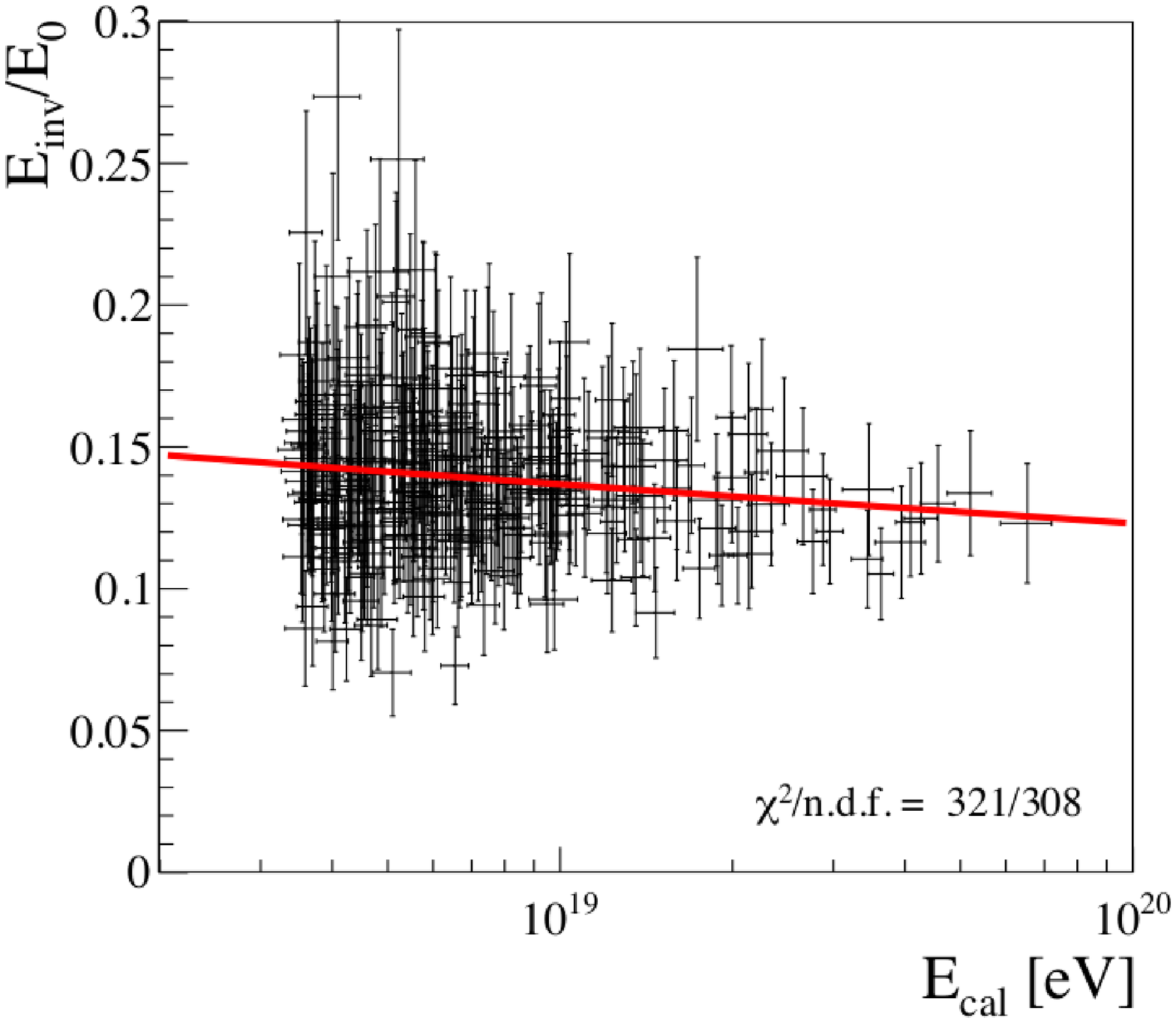}
\includegraphics[width=0.9\columnwidth,height=0.9\columnwidth,keepaspectratio,clip]{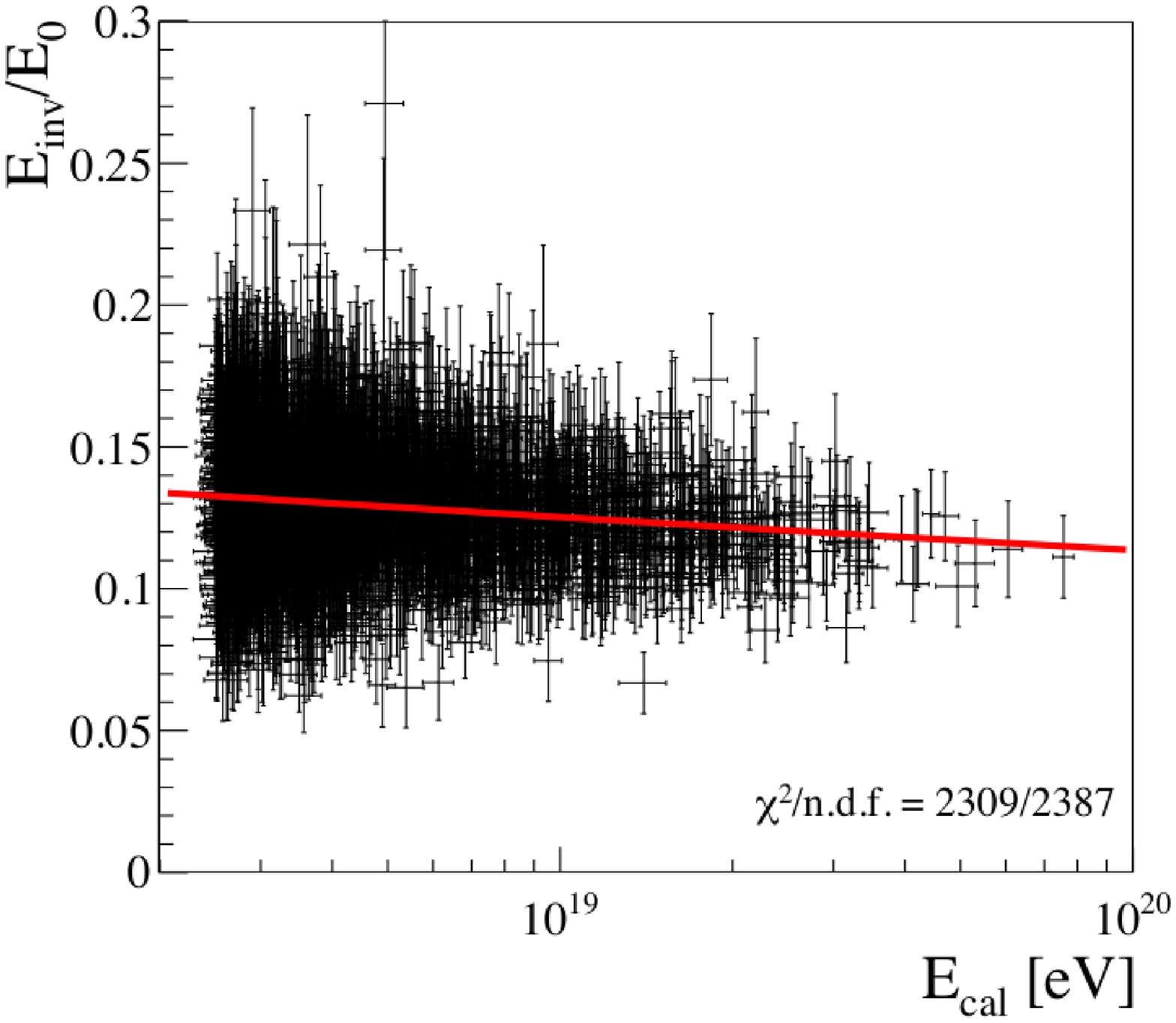}
 }
\caption{$\frac{E_{\rm inv}}{E_0}$ as a function of $E_{\rm cal}$ from the inclined (left) and vertical (right) hybrid events. Fitted function shown with red line.}
\label{fig:EinvData}
\end{figure*}
\subsection{\texorpdfstring{$E_{\rm inv}$)}{ei} from vertical showers}
\label{Sec:Einv_S1000}
As seen in Sec.~\ref{Sec:EinvPhenomenology}, $\einv$ 
is a power law function of $E_0$ 
\begin{equation}
E_{\rm inv} =  \epsilon_c^{\pi}  \beta_0  \left(\frac{E_0}{\epsilon_c^{\pi}}\right)^\beta.
\label{eq:Einv_E0_beta0}
\end{equation}
$\beta_0$, equal to $A^{1-\beta}$ in the extended Heitler model~\cite{Matthews} (see Eq.~\eqref{eq:EinvE0A}), is a parameter introduced in order to account for the large variations in the predictions of the number of muons that are obtained using different hadronic interaction models once $E_0$ and primary mass are fixed.

In the reconstruction of vertical events, $E_0$ is estimated from $S(1000)$, the signal at $1000$ m from the core~\cite{Auger}, by correcting for the shower attenuation using the constant intensity cut method~\cite{CIC}. To estimate $E_{\rm inv}$ from $S(1000)$, we use the functional form
\begin{equation}
 E_{0}= \gamma_0(\Delta X)\,\left[S(1000)\right]^{\gamma},
\label{eq:E0_CIC}
\end{equation}
where $\Delta X = X - X_{\rm max}$ is the atmospheric slant depth between ground level at the Auger site and the depth of the shower maximum development, 
and $\gamma_0(\Delta X)$ is related to the attenuation of $S(1000)$ with $\Delta X$. 

Combining Eq.~\eqref{eq:Einv_E0_beta0} and \eqref{eq:E0_CIC} one obtains 
\begin{eqnarray}
E_{\rm inv} & = & \epsilon^{\pi}_{c} ~\beta_0 ~\left(\frac{\gamma_0(\Delta X)\,S(1000)^{\gamma}} {\epsilon^{\pi}_{c}}\right)^{\beta} \\
 & = & A(\Delta X) ~\left[S(1000)\right]^{B},
\label{eq:Einv_S1000}
\end{eqnarray}
where 
\begin{eqnarray}
\label{eq:A_DX}
A(\Delta X) & = &  \left( \epsilon^{\pi}_{c} \right)^{1-\beta}   ~\beta_0 ~\left[ \gamma_0(\Delta X) \right]^\beta, \\
B & = & \gamma \beta~.
\label{eq:B}
\end{eqnarray}
The parameter $B$ and those defining the function $A(\Delta X)$ are determined using Monte Carlo simulations. 
Using the \qgsjet II-03 hadronic interaction model, we find $\beta=0.925$ and $\gamma=1.0594$, so that their product is $B=0.98$. Different interaction models yield the same value of $B$ to within 2\%. This value will be used from now on, so that with Eq.~\eqref{eq:Einv_S1000} and the measurements of $S(1000)$ and $\Delta X$ one can obtain an event-by-event estimate of $E_{\rm inv}$.
The function $A(\Delta X)$ is calculated using events simulated with the \qgsjet II-03 hadronic interaction model for a mixed composition of 50\% protons and 50\% iron. $A(\Delta X)$ is parameterised with the fourth-degree polynomial.

The performance of the analysis is tested with proton and iron events simulated with the hadronic interaction models \qgsjet II-03, and EPOS 1.99.
The average value of the residuals as a function of $E_0$, shown in~Fig.~\ref{fig:EinvS1000Bias} (left), are between $-5 \%$ and $20 \%$. The spread in the residuals is mainly due to the difference in the predictions of the number of muons and of the  attenuation function $\gamma_0(\Delta X)$ among the simulations used to parametrise $A(\Delta X)$, and the ones used to simulate the events. 
Note that the function $\gamma_0(\Delta X)$ includes the conversion factor needed to obtain $E_0$ from $S(1000)$ which is strongly model dependent.

A better estimation of $\einv$ can be obtained taking into account these differences using the following equation
\begin{equation}
 E_{\rm inv} = A(\Delta X) ~\left[S(1000)\right]^{B} ~\left(   \frac{ \tilde{\gamma}_0(\Delta X)}{    \gamma_0(\Delta X)  }  \right)^\beta  ~\frac{\tilde{\beta}_0}{\beta_0},
\label{eq:correction}
\end{equation}
where the quantities with and without the accent tilde are calculated for the data sample that we are analysing and for the one used to parametrise $A(\Delta X)$, respectively. $\beta$ is fixed to 0.925. 
The functions $\gamma_0$ are obtained from Eq.~\eqref{eq:E0_CIC} using $E_0$ and $S(1000)$. The ratio $\tilde{\beta}_0/\beta_0$ is estimated from the ratio of the number of muons at ground level for the two data sets, information that is available in the CORSIKA events. 
The residuals in $\einv$ using the improved parameterisation of Eq.~\eqref{eq:correction} are shown in Fig.~\ref{fig:EinvS1000Bias} (right). 
The true value of $E_{\rm inv}$ can be recovered within a few \% for all models and primaries. 
Note also how we improve the estimation of $\einv$ for \qgsjet II-03, despite the primary mass composition used to parametrise $A(\Delta X)$ being different to that of the simulated events used to test the analysis method. 
%
\begin{figure*}[t]
\centering
 \centerline{
   \includegraphics[width=1\columnwidth,height=1\columnwidth,keepaspectratio,clip] {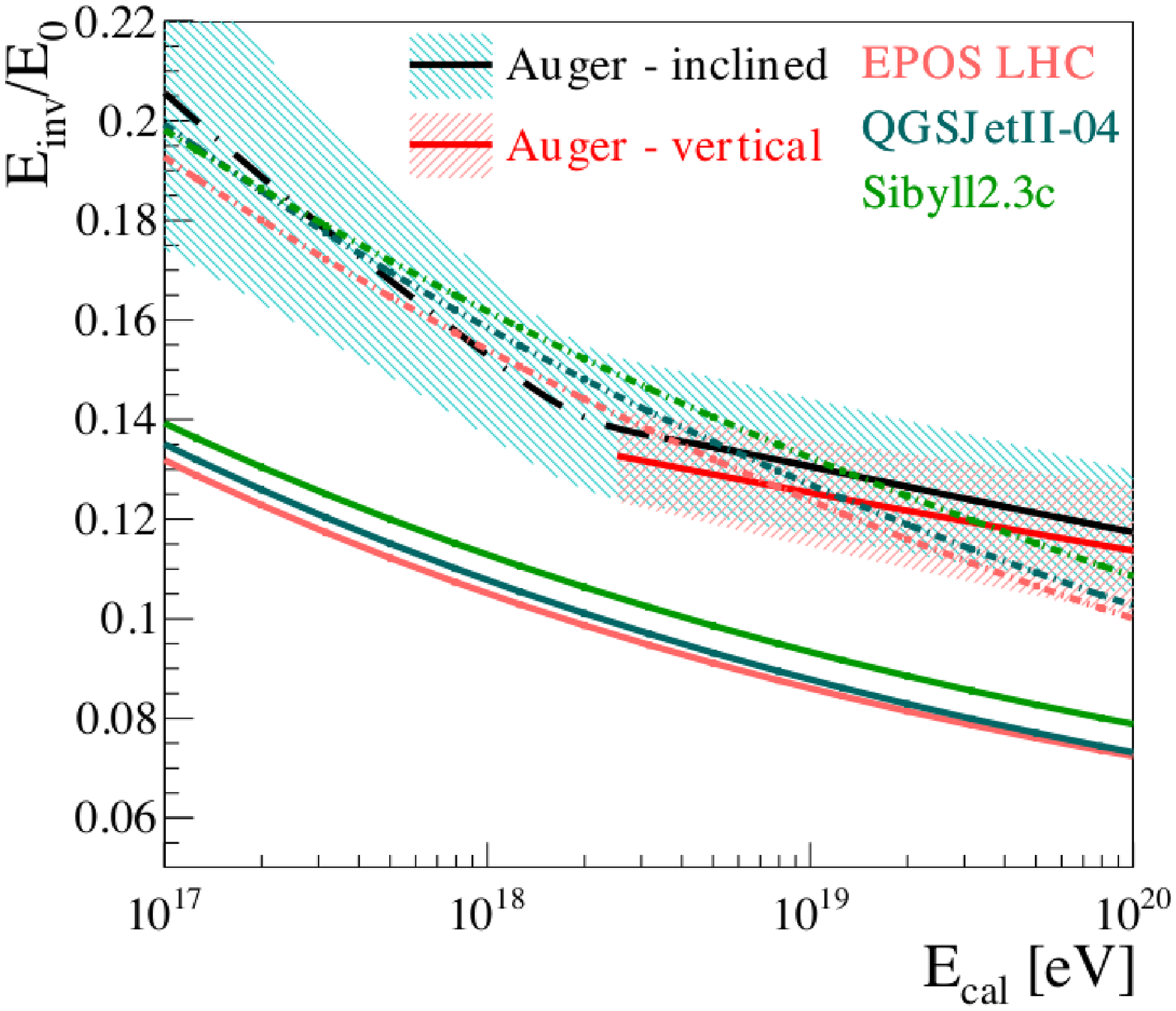}
   \includegraphics[width=1\columnwidth,height=1\columnwidth,keepaspectratio,clip]{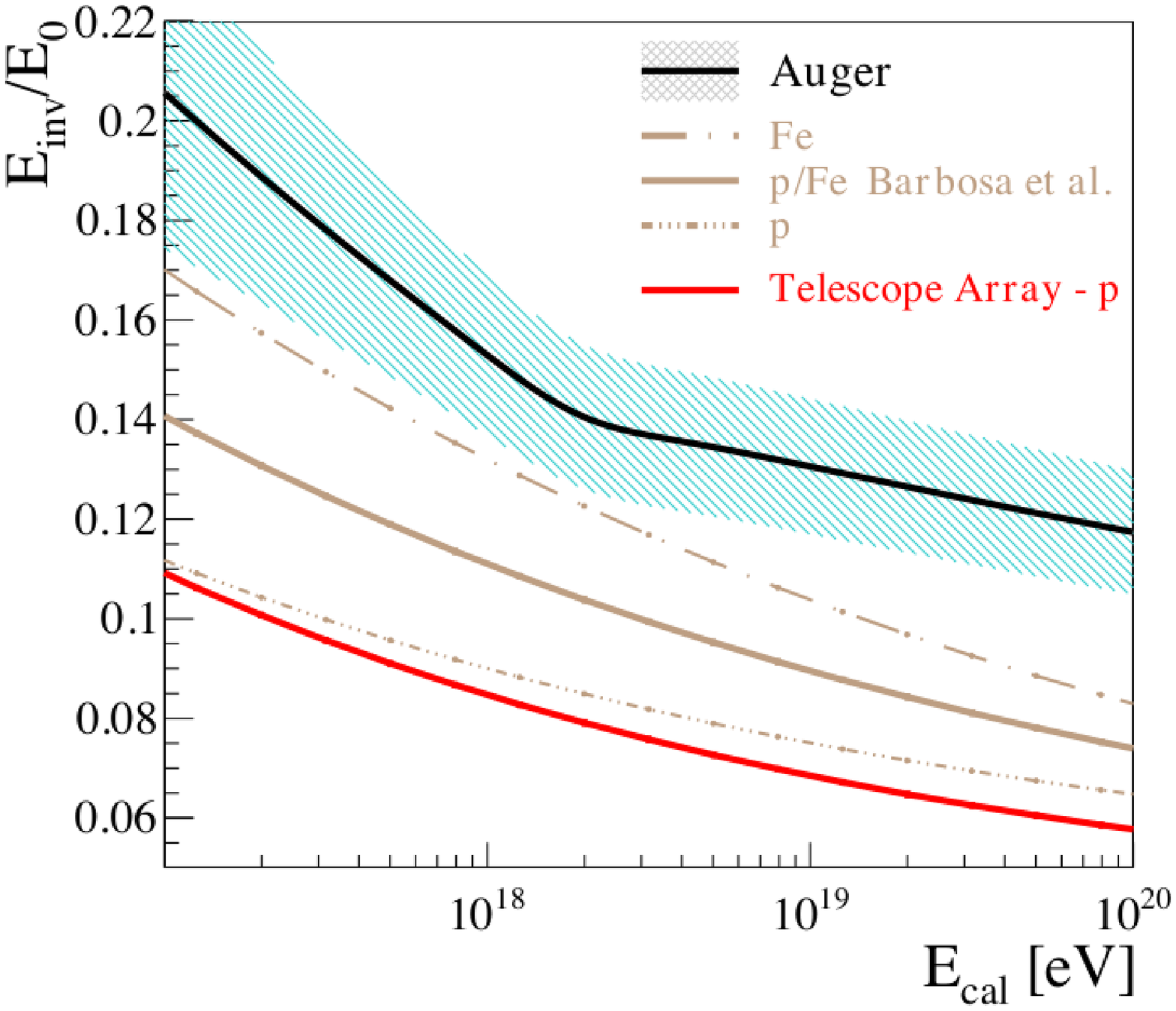}
           }
\caption{Left:$E_{\rm inv}$ for inclined and vertical events compared with predictions from simulations. Systematic uncertainty are shown with the shaded bands. The estimate for inclined events is extrapolated to low energies. 
Right:Auger data-driven estimation of $E_{\rm inv}$ compared with the parameterisations for protons, iron and mixed composition reported in~\cite{Barbosa} and the one in use by Telescope Array~\cite{TA-Einv}. 
          }
\label{fig:Einv-ecal}
\end{figure*}
\subsection{Parameterisation of \texorpdfstring{$E_{\rm inv}$}{ei} as a function of \texorpdfstring{$E_{\rm cal}$}{ec} } 
\label{Sec:Einv_data}
The analysis methods described in Sec.~\ref{Sec:Einv_Rmu} and~\ref{Sec:Einv_S1000} allow us to obtain an event-by-event estimation of $E_{\rm inv}$ from the data collected by the Pierre Auger Observatory. 

The analysis is limited to events sufficiently energetic to ensure a full SD trigger efficiency. At energies lower than $4 \times 10^{18}$ eV for the inclined~\cite{HASRec} and $3 \times 10^{18}$ eV for the vertical events~\cite{SDExp}, the trigger is biased towards events with a higher number of muons, and thus higher $E_{\rm inv}$ and consequently larger systematic uncertainties. 
In order to get an estimation of $\einv$ useful for all FD events, including the ones with energies below the full SD trigger efficiency, the event-by-event estimation of $E_{\rm inv}$ is parameterised as a function of $E_{\rm cal}$ above the full trigger efficiency, with the function being extrapolated to lower energies.

Analysing hybrid events collected from 1 January 2004 to 31 December 2015, selected with the same cuts used for the energy calibration of the SD energy estimators~\cite{Auger-EnSc} a parameterisation was obtained.

The correlation between $E_{\rm inv}$ and $E_{\rm cal}$ is well approximated by a power law relationship
\begin{equation}
E_{\rm inv} = a \left( \frac{E_{\rm cal}}{10^{18} {\rm~ eV} } \right)^b.
\label{eq:Einv_Ecal}
\end{equation}
The data and the fitted function are shown in Fig.~\ref{fig:EinvData}. 

For a quantitative comparison of the two data-driven estimations of $\einv$ one has to take into account $\einv$ zenith angle dependence. Since the majority of the events are below $60^\circ$, 
the $E_{\rm inv}$ parameterisation from the inclined data set, that is on average 5\% than the vertical one, has been corrected.
The two data-driven $\einv$ estimations are compared in Fig.~\ref{fig:Einv-ecal} (left) together with the theoretical predictions for post-LHC hadronic interaction models. They are still larger than the predictions for iron primaries, in contradiction with the mean mass obtained using $X_{\rm max}$ measurements~\cite{Auger-mass}. 
This is due to the muon deficit~\cite{AugerMuonSize} as models fail to describe the properties of shower development related to muons and therefore to $E_{\rm inv}$.

It is worth noting that the two estimates are partially correlated since they both use the measurement of $N_\mu$.
However, they are affected by different systematics. 

The estimations of $\einv$ obtained above the energy of full SD trigger efficiency can be extrapolated to lower energies taking into account the change in the mean mass composition evolution with energy at $E_{\rm cal}^A \simeq 2 \times 10^{18} \rm{eV}$ measured by Auger~\cite{Auger-mass,Auger-mass-comp-ICRC17}. 
The function is obtained by extrapolating the parameterisation obtained from data down to $E_{\rm cal}^A$ and, below this energy, using a model inspired function that matches the parameterisation at $E_{\rm cal}^A$. For the latter, we use the function of Eq.~\eqref{eq:EinvE0A} in which the mean composition as a function of energy is taken from the Auger FD measurements~\cite{Auger-mass-comp-ICRC17} together with a value of $\beta = 0.9$ that reproduces the simulations at lower energies. 
The extrapolation of $\einv$ obtained from the inclined events, shown with the black dashed line in Fig.~\ref{fig:Einv-ecal} (left), will be replaced in the near future with a more accurate estimation of $\einv$ using the data collected by the AMIGA muon detectors~\cite{AMIGA} installed at the Observatory and using the 750m-spacing sub-array of WCDs~\cite{Auger}.
\section{Conclusions}
A data driven estimation of $E_{\rm inv}$ of cosmic ray showers detected by the Pierre Auger Observatory, was presented. Two analysis methods for inclined ($60^\circ < \theta < 80^\circ$) and vertical ($\theta < 60^\circ$) events were developed.
$E_{\rm inv}$ has been parameterised as a function of $E_{\rm cal}$ and extrapolated to energies below the SD full trigger efficiency. The two estimations agree at a level well within the systematic uncertainties, that are estimated to be of the order of $10\%-15\%$.

$E_{\rm inv}$ results are considerably higher than the predictions given by simulations. This is a consequence of the muon deficit in models~\cite{AugerMuonSize}, a deficit due to the failure of the hadronic interaction models to describe the properties of shower development related to muons. 
Moreover, the results are consistent with the evolution of the mass with energy as measured by Auger~\cite{Auger-mass,Auger-mass-comp-ICRC17}. This is due to the sensitivity of $N_\mu$ to the primary mass and, at lower energy, due to the use of the mean mass composition to find the functional form that describes $\einv$ as a function of $\ecal$.

The measurement of $N_\mu$ makes the analysis of $\einv$ from inclined showers rather straightforward and intrinsically better than the analysis used for vertical events.


\end{document}